\documentclass{article}
\usepackage[utf8]{inputenc}
\usepackage{graphicx}
\usepackage{hyperref}
\usepackage{xcolor} 
\usepackage{amsmath} 
\usepackage[percent]{overpic} 
\usepackage{subcaption}  
\usepackage[normalem]{ulem} 
\usepackage{authblk}

\usepackage[style = ieee, citestyle=numeric, backend=biber, sorting=none, dashed=false, eprint=false, isbn=false]{biblatex}
\AtEveryBibitem{\clearlist{language}}
\AtEveryBibitem{\clearfield{month}}
\AtEveryBibitem{%
  \ifboolexpr
    {
      test { \ifentrytype{book} }
      or
      test { \ifentrytype{article} }
      or
      test { \ifentrytype{inproceedings} }
    }
    {\clearfield{url}}
    {}%
}
\AtEveryBibitem{
    \clearfield{urlyear}
    \clearfield{urlmonth}
}

\newcommand{\vect}[1]{\boldsymbol{#1}} 

\title{Ferromagnetic filament shapes in a rotating field reveal their magnetoelastic properties}
\author[1]{A. P. Stikuts\footnote{andris\_pavils.stikuts@lu.lv}}
\author[1]{A. C\={e}bers\footnote{andrejs.cebers@lu.lv}}
\author[1]{G. Kitenbergs\footnote{guntars.kitenbergs@lu.lv}}
\affil[1]{MMML lab, Department of Physics, University of Latvia, Jelgavas 3, Riga, LV-1004, Latvia}
\date{June 2023}

\begin{document}

\maketitle

\section*{Abstract}
Flexible ferromagnetic filaments can be used to control the flow on the micro-scale with external magnetic field. 
To accurately model them, it is crucial to know their parameters such as their magnetization and bending modulus, the latter of which is hard to determine precisely. 
We present a method how the ferromagnetic filament's shape in a rotating field can be used to determine the magnetoelastic number $Cm$ - the ratio of magnetic to elastic forces.
Then once the magnetization of the filament is known, it is possible to determine its bending modulus.
The main idea of the method is that $Cm$ is the only parameter that determines whether the filament is straight or whether its tips are bent towards the magnetic field direction.
Comparing with numerical solutions, we show that the method results in an error of $15...20\%$ for the determined $Cm$, what is more precise than estimations from other methods.
This method will allow to improve the comparability between theoretical filament models and experimental measurements.

\section{Introduction}

Magnetic filaments can be created by connecting paramagnetic or ferromagnetic beads with a linker, for example, DNA fragments or some other polymer
\cite{goubault_flexible_2003,biswal_rotational_2004,yang_superparamagnetic_2018,spatafora-salazar_hierarchical_2021}. 
The resulting filaments are typically tens of microns long.
The diameter of the beads (and thus the width of the filament) range from less than a micron for the paramagnetic case \cite{goubault_flexible_2003} to $4~\mu$m for the ferromagnetic case \cite{erglis_three_2010}.
Apart from the choice of the beads, another variable that determines the properties of the filaments is the choice and length of the linker polymers. 
The impact of DNA linker length to the filament's flexibility was analyzed in ref. \cite{byrom_directing_2014}.
Magnetic filaments can be used to influence the flow on the micro-scale using external magnetic fields.
A few of their applications include microswimming \cite{dreyfus_microscopic_2005,zaben_instability_2021}, micromixing \cite{biswal_micromixing_2004, zhang_metachronal_2022}, and
navigation through microfluidic channels \cite{yang_reconfigurable_2020}.

To describe such filaments theoretically \cite{cebers_flexible_2016}, it is vital to accurately determine their magnetization $M$ and bending modulus $A_B$ (such that the associated energy is $A_B\int k^2 dl/2  $, where $k$ is the curvature of the filament).
The magnetization depending on the applied magnetic field can be determined from bulk measurements of the beads that make up the filaments \cite{erglis_three_2010}.
The determination of the bending modulus $A_B$ requires more subtle techniques since the filaments are micron-sized.
Paramagnetic filaments when placed in a static field form long lived metastable hairpin-like U shapes, whose maximum curvature $k_{max}$ for strong fields is proportional to the square root of the magnetoelastic number $Cm$ (the ratio of magnetic to elastic forces) \cite{cebers_flexible_2016, belovs_equilibrium_2022}.
From this it is possible to determine the value of $A_B$ \cite{goubault_flexible_2003}.
Theoretically stationary $U$ shapes can also be achieved with ferromagnetic filaments, however they are more unstable than the paramagnetic U shapes, and quickly relax to straight shapes through the third dimension.
Nonetheless it has been attempted to estimate $A_B$ by observing the curvature just before the relaxation \cite{erglis_three_2010}.
Another approach uses the fact that a slightly deformed filament exponentially relaxes to a straight shape with the characteristic time 
$t_0=\zeta_\perp L^4 /A_B$, where $\zeta_\perp$ is the drag coefficient in the direction normal to the filament's centerline \cite{wiggins_trapping_1998}.
Using this method, $A_B$ was estimated in ref. \cite{zaben_deformation_2020-1}, where the drag coefficient was estimated $\zeta_\perp\approx 4\pi \eta$ with $\eta$ being the viscosity of the surrounding fluid.  
This, however, might be an underestimate since the filaments are close to the bottom of the sample cell and the drag close to it rapidly increases \cite{koens_local_2021}.

Many experimental works \cite{erglis_three_2010,zaben_instability_2021,zaben_deformation_2020-1, erglis_flexible_2009,zaben_3d_2020} use ferromagnetic filaments formed using streptavidin coated micron sized ($d=4.26 \mu m $) ferromagnetic beads (Spherotec, 1\%w/v) that are linked with 1000bp long biotinized DNA fragments (ASLA biotech or Latvian Biomedical Research and Study Centre) following the procedure outlined in \cite{erglis_experimental_2010}.
The estimated bending modulus $A_B$ values for some of these works are gathered in table \ref{tab:measured_AB}. 
There is a variation of several orders of magnitude, which motivates us to devise a relatively simple procedure to determine the filament's bending modulus.

\begin{table}[h]
    \centering
    \begin{tabular}{c|c}
        Method & $A_B, J\cdot m$ \\ \hline
        U shape curvature estimation \cite{erglis_three_2010} & $1.5\cdot10^{-19}$ \\
        Relaxation to a straight shape\footnotemark \cite{zaben_deformation_2020-1} & $4.1\cdot10^{-23}$ \\ 
        Fitting of numerical 3D precessing shape \cite{zaben_3d_2020} & $1.5\cdot10^{-21}$
    \end{tabular}
    \caption{The determined bending modulus $A_B$ of several experimental works that use the same type of ferromagnetic filaments.}
    \label{tab:measured_AB}
\end{table}
\footnotetext{We have corrected the final step of the calculation, where $L$ in formula $3.93^4L^{-4} A_B/\zeta$ needs to be taken as \textit{half} of the filament's length.}
In this work we show how a ferromagnetic filament's shape in a rotating field can be used to determine the magnetoelastic number $Cm$ and thus the bending modulus $A_B$. 
We solve for the equilibrium shape of the filament for small deviation from the magnetic field direction. 
We then extend this solution to include large deviations when $Cm$ is small.
Finally we outline the procedure how to determine the filament's parameters and compare it to full numerical simulations to determine its accuracy.

\section{Mathematical model}
\label{sec:math_model}
Elastic magnetic filaments are commonly modeled using Kirchhoff theory of an elastic rod with additional terms that describe the magnetic interactions \cite{cebers_flexible_2016,cebers_dynamics_2003}.
The filament of length $L$ is described by the radius vector $\vect r(l)$ which is parameterized by the arc length $l\in[-L/2,L/2]$.
The force acting on the cross-section of the filament reads
\begin{equation}
    \vect F = -A_B \vect r_{lll}+\Lambda \vect r_l + \vect F_m,
\end{equation}
where the subscript $l$ denotes the derivative with respect to the arc length, $A_B$ is the bending modulus and $\Lambda(l)\vect r_l$ is the tension force that ensures the inextensibility of the filament.
$\vect F_m$ is the magnetic force that for a ferromagnetic filament reads
\begin{equation}
    \vect F_m = -\mu_0M\vect H,
\end{equation}
where $M$ is the magnetic moment per unit length of the filament, and $\vect H$ is the applied magnetic field intensity.

When the filament is slender, its motion can be described by the resistive-force theory \cite{lauga_fluid_2020}. 
The linear force density in a Stokes flow is connected with the velocity through the drag coefficients parallel to the filament $\zeta_\parallel$ and perpendicular to it $\zeta_\perp$
\begin{equation}
    \vect F_l = \zeta_\perp ((\vect v - \vect v_\infty)\cdot \vect n) \vect n + \zeta_\parallel ((\vect v-\vect v_\infty)\cdot \vect t)\vect t,
\end{equation}
where $\vect t$ and $\vect n$ is the tangent and normal vectors of the filament, respectively, and $\vect v_\infty$ is the background flow velocity.
The local inextensibility of the filament dictates that
\begin{equation}
    \vect r_l \cdot \vect r_l = 1.
\label{eq:inext}
\end{equation}
Taking the time derivative of equation \eqref{eq:inext} we get a constraint on the velocity $\vect r_l \cdot \vect v_l = 0$.

Finally, the mathematical model is concluded with the boundary conditions of torque and force free filament ends, which at $l=\pm L/2$ require
\begin{equation}
    \vect r_{ll} |_{l=\pm L/2} = \vect 0
\end{equation}
and
\begin{equation}
    (-A_B\vect r_{lll} + \Lambda \vect r_l -\mu_0 M \vect H)|_{l=\pm L/2} = \vect 0.
\end{equation}

\subsection{Dimensionless parameters}
The mathematical model can be rendered dimensionless by introducing the following scales:
\begin{itemize}
    \item length scale $r_0=L$,
    \item time scale $t_0=\zeta_\perp L^4 /A_B$.
\end{itemize}
With this scaling, dimensionless parameters appear in the mathematical formulation:
\begin{itemize}
    \item the magnetoelastic number $Cm=\mu_0 M H L^2 / A_B$,
    \item the ratio of perpendicular and parallel drag coefficients $\zeta_\perp/\zeta_\parallel$.
\end{itemize}
The ratio of drag coefficients is close to 2 for slender filaments \cite{lauga_fluid_2020}.
If there is a rotating magnetic field driving the filament, a third dimensionless parameter arises:
\begin{itemize}
    \item the Mason number $M_a=\omega \zeta_\perp L^2 / (12 \mu_0 M H)$,
\end{itemize}
where $\omega$ is the angular frequency of the field.
The equations rendered dimensionless by $r_0$ and $t_0$ are used in section \ref{sec:derivation}, where the equilibrium shape for small deformations is derived. 
Elsewhere to facilitate the reading, dimensional formulas are used.

\section{Derivation of the equilibrium shape of a filament in a rotating field}
\label{sec:derivation}

\begin{figure}
    \centering
    \includegraphics[width=0.7\textwidth]{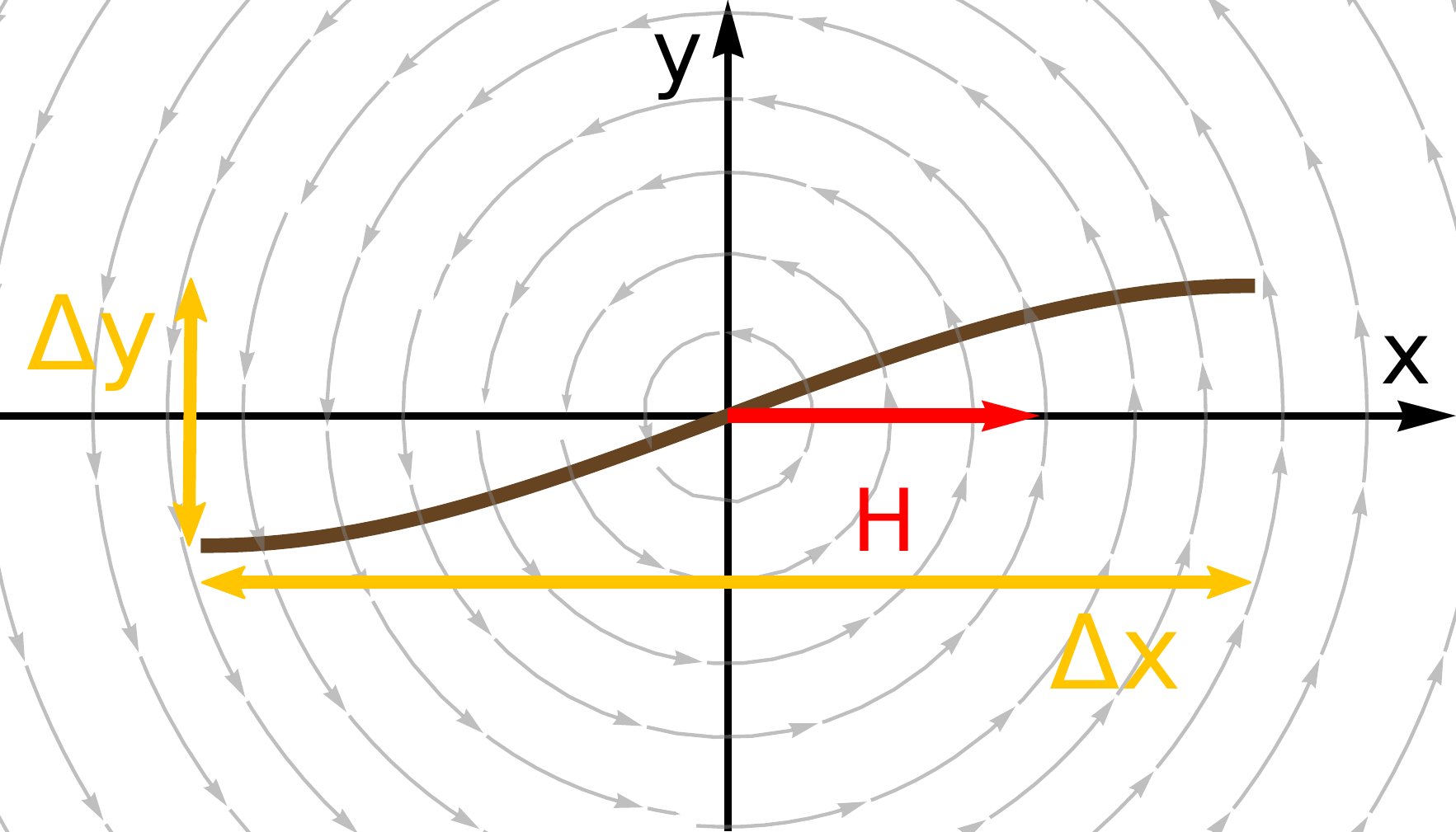}
    \caption{Sketch of the magnetic filament equilibrium shape in the reference frame of the rotating magnetic field. $\Delta x$ and $\Delta y$ show the difference of coordinates between the tips of the filament. The field is rotating clockwise, therefore in this reference frame there is a background flow counter-clockwise.}
    \label{fig:rotating_system}
\end{figure}

To find the equilibrium shape of the filament in a rotating magnetic field, we can utilize the lack of inertia in the Stokes flow regime, and move to a coordinate system that rotates with the magnetic field (figure \ref{fig:rotating_system}).
We set the magnetic field along the $x$ axis.
A background flow of $\vect v_\infty=\{-\omega y, \omega x, 0\}$ arises, where $\omega$ is the angular velocity of the magnetic field.
When $\omega=0$, the filament lies along the x axis. 
We seek an approximate equilibrium shape up to first order in $\omega$ and $y$.
The arclength parameter becomes $l=x+O(y^2)$.

With this approximation, along the $x$ axis the equations stated in the section \ref{sec:math_model} read
\begin{equation}
    \Lambda_x = 0,
\end{equation}
where the subscript $x$ denotes the derivative with respect to $x$.
This together with the only non-trivial boundary condition
\begin{equation}
    (-Cm + \Lambda)|_{x = \pm 1/2} = 0
\end{equation}
gives us the solution for the tension force
\begin{equation}
    \Lambda=Cm.
\label{eq:Lambda}
\end{equation}

Along the $y$ axis the equations stated in the section \ref{sec:math_model} read
\begin{equation}
    -y_{xxxx} + \Lambda y_{xx} +\omega x = 0.
\end{equation}
The boundary conditions are
\begin{equation}
    y_{xx}|_{x=\pm 1/2}=0,
\end{equation}
\begin{equation}
    (-y_{xxx}+\Lambda y_x)|_{x=\pm 1/2}=0.
\end{equation}
Plugging in $\Lambda$ from equation \eqref{eq:Lambda}, and requiring that $y(0)=0$, we get the solution for the equilibrium shape of the filament
\begin{equation}
    y=\frac{\omega}{12 Cm} \left[ \frac{6\sinh{(x\sqrt{Cm})}}{Cm\sinh{(\sqrt{Cm}/2)}} 
    -2x^3 + x \left(\frac{3}{2} -\frac{12}{Cm} \right)
    \right].
\label{eq:eq_shape}
\end{equation}

It is possible to verify that up to the first order in $y$, the filament inextensibility condition (eq. \eqref{eq:inext}) is satisfied.
Additionally from the equation \eqref{eq:eq_shape} it is evident that $y$ is small when $\omega/(12Cm)$ is small, which suggests that $\omega/(12Cm) \ll 1$ is the criterion for the validity of this solution.
The square-bracketed expression is bounded between $\pm 1/2$ for $x\in[-1/2,1/2]$.

\section{Analysis of the asymptotic equilibrium shape}

For convenience from now on we will again use dimensional expressions.
Equation \eqref{eq:eq_shape} for the equilibrium shape in dimensional form reads
\begin{equation}
    \frac{y}{L}=M_a \left[ \frac{6\sinh{(x\sqrt{Cm}/L)}}{Cm\sinh{(\sqrt{Cm}/2)}} 
    -2\frac{x^3}{L^3} + \frac{x}{L} \left(\frac{3}{2} -\frac{12}{Cm} \right)
    \right],
\label{eq:dim_shape}
\end{equation}
where we identify the coefficient in front of of the brackets as the Mason number $M_a=\omega \zeta_\perp L^2 / (12 \mu_0 M H)$,
which is the ratio of viscous to magnetic forces in the system. 

\begin{figure}[h]
    \centering
    \includegraphics[width=0.7\textwidth]{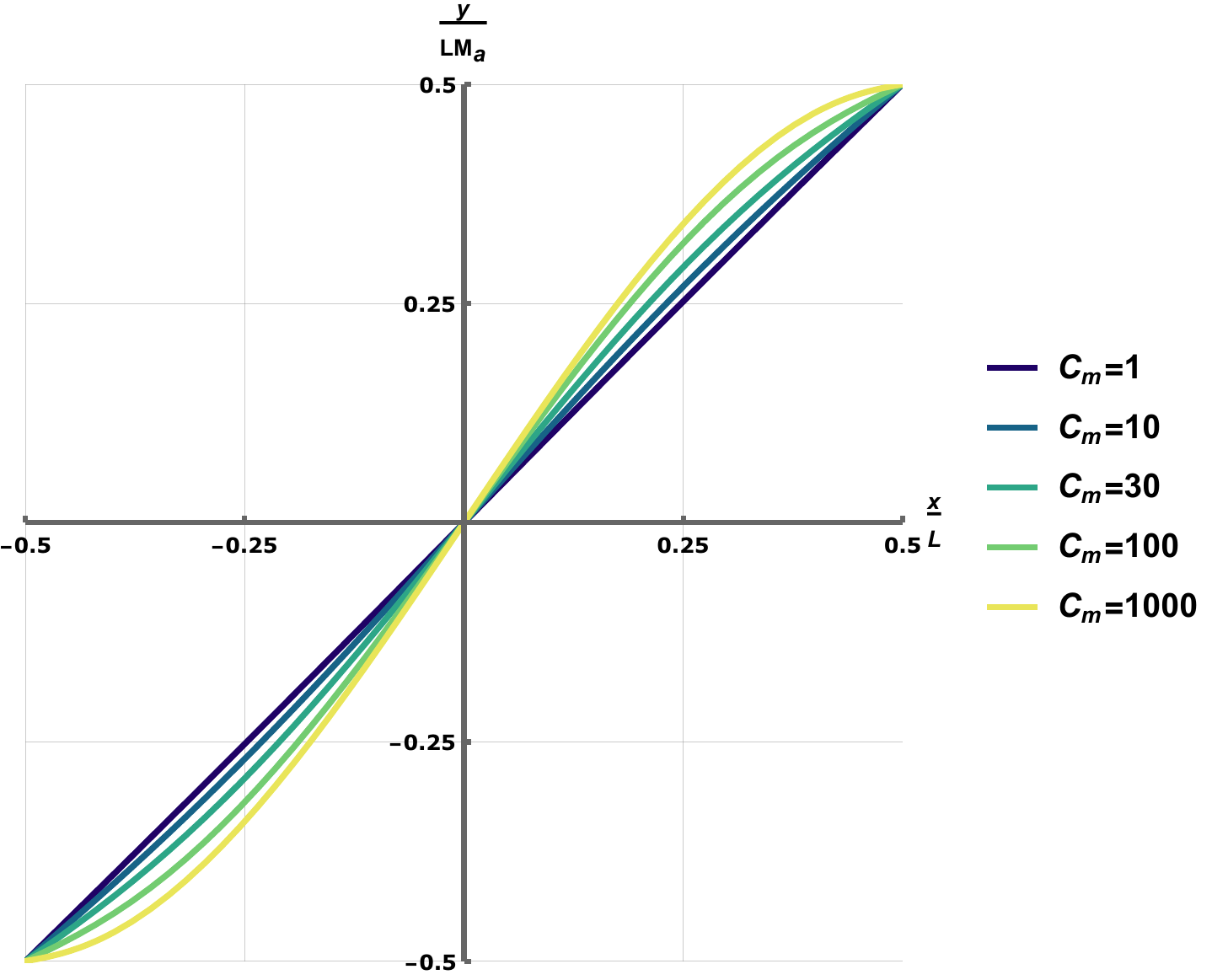}
    \caption{Illustration of how the magnetoelastic number $Cm$ governs the shape of the filament according to eq. \eqref{eq:dim_shape}. }
    \label{fig:shape}
\end{figure}

Interestingly, the tip coordinates of the filament are independent of $Cm$ and are determined solely by $M_a$. 
Denoting with $\Delta x$ and $\Delta y$ the difference of $x$ and $y$ coordinates between the tips (figure \ref{fig:rotating_system}), we can write
\begin{equation}
    \frac{\Delta y}{ \Delta x} = M_a.
    \label{eq:Ma_eq}
\end{equation}
This means that the expression in the square brackets solely determines the shape of the filament connecting the two end points. 
We plot the equation \eqref{eq:dim_shape} divided with $M_a$ for different values of $Cm$ (see figure \ref{fig:shape}) to observe how this happens.
For small values of $Cm$, the filament is nearly straight (which is expected, since in the limit of $A_B\rightarrow \infty $, the filament should be a rigid rod).
Whereas large $Cm$ means that the tips of the filament become bent in the direction of the field.

\section{Correction to the asymptotic solution to take into account the limit of a rigid rotating rod }

It is possible to write the equation for the deviation of a rigid ferromagnetic rod from the magnetic field direction \cite{goyeau_dynamics_2017}
\begin{equation}
    \sin{\theta} = \frac{\Delta y}{L} = M_a,
\label{eq:Ma_eq_l}
\end{equation}
where $\theta$ is the angle between the rod and the field, $\Delta y$ and $L$ are defined the same way as for the flexible filament.
In terms of $y(l)$ the equation for the rigid rod ($Cm\rightarrow0 $) reads
\begin{equation}
    y=M_a l .
\label{eq:rod_shape}
\end{equation}
Note that unlike eq. \eqref{eq:dim_shape}, eq. \eqref{eq:rod_shape} is valid for arbitrary deviations from the magnetic field direction.

Knowing this, we can modify eq. \eqref{eq:dim_shape} to include the limit of rigid rod as $Cm\rightarrow 0$.
We write 
\begin{equation}
    \frac{y}{L}= M_a \left[ \frac{6\sinh{(l\sqrt{Cm}/L)}}{Cm\sinh{(\sqrt{Cm}/2)}} 
    -2\frac{l^3}{L^3} + \frac{l}{L} \left(\frac{3}{2} -\frac{12}{Cm} \right)
    \right],
\label{eq:dim_shape_l}
\end{equation}
where we replaced $x \rightarrow l$.
For small $y/L$, this corrected expression is asymptotically identical to the previously derived eq. \eqref{eq:dim_shape}.
Whereas in the limit of small $Cm$, it is identical to the rigid rod (eq. \eqref{eq:rod_shape}) for arbitrary $y/L$.
Note that the formula only gives the $y$ coordinate of the filament, however, the $x$ coordinate can be determined by integrating $dx/dy = \sqrt{(dl/dy)^2 -1}$.

\begin{figure}[h]
    \centering
    \includegraphics[width=0.32\textwidth]{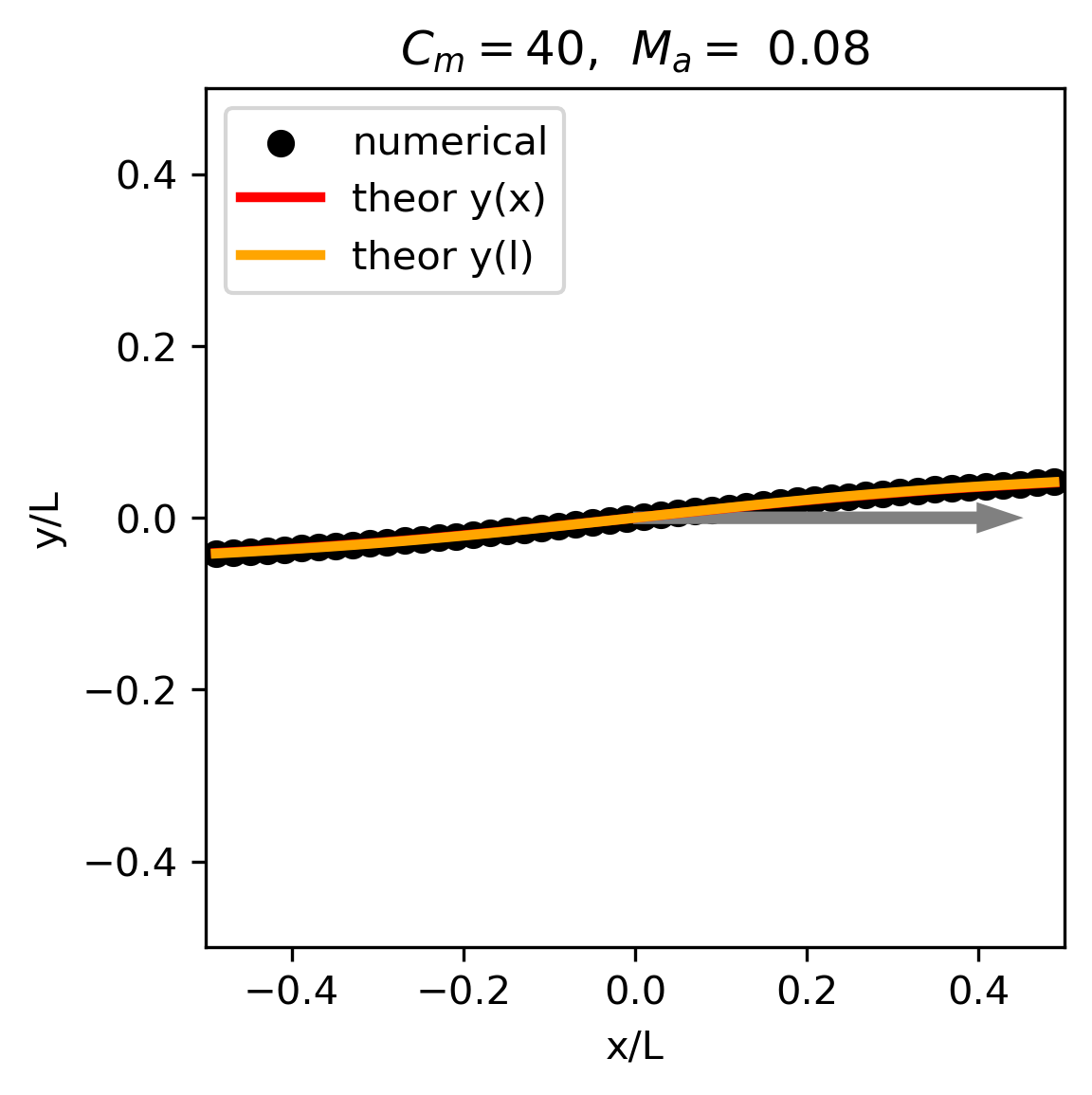}
    \includegraphics[width=0.32\textwidth]{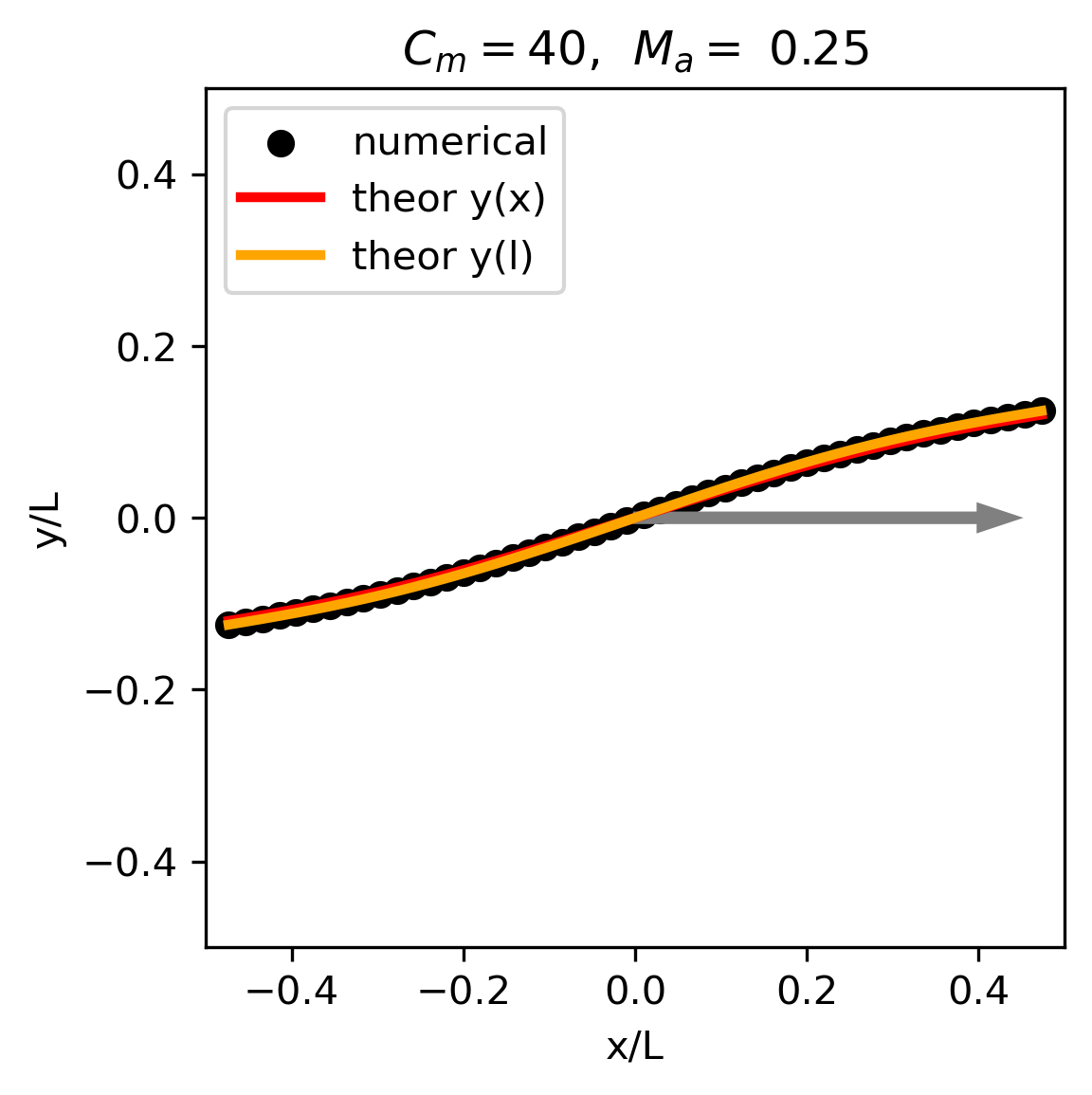}
    \includegraphics[width=0.32\textwidth]{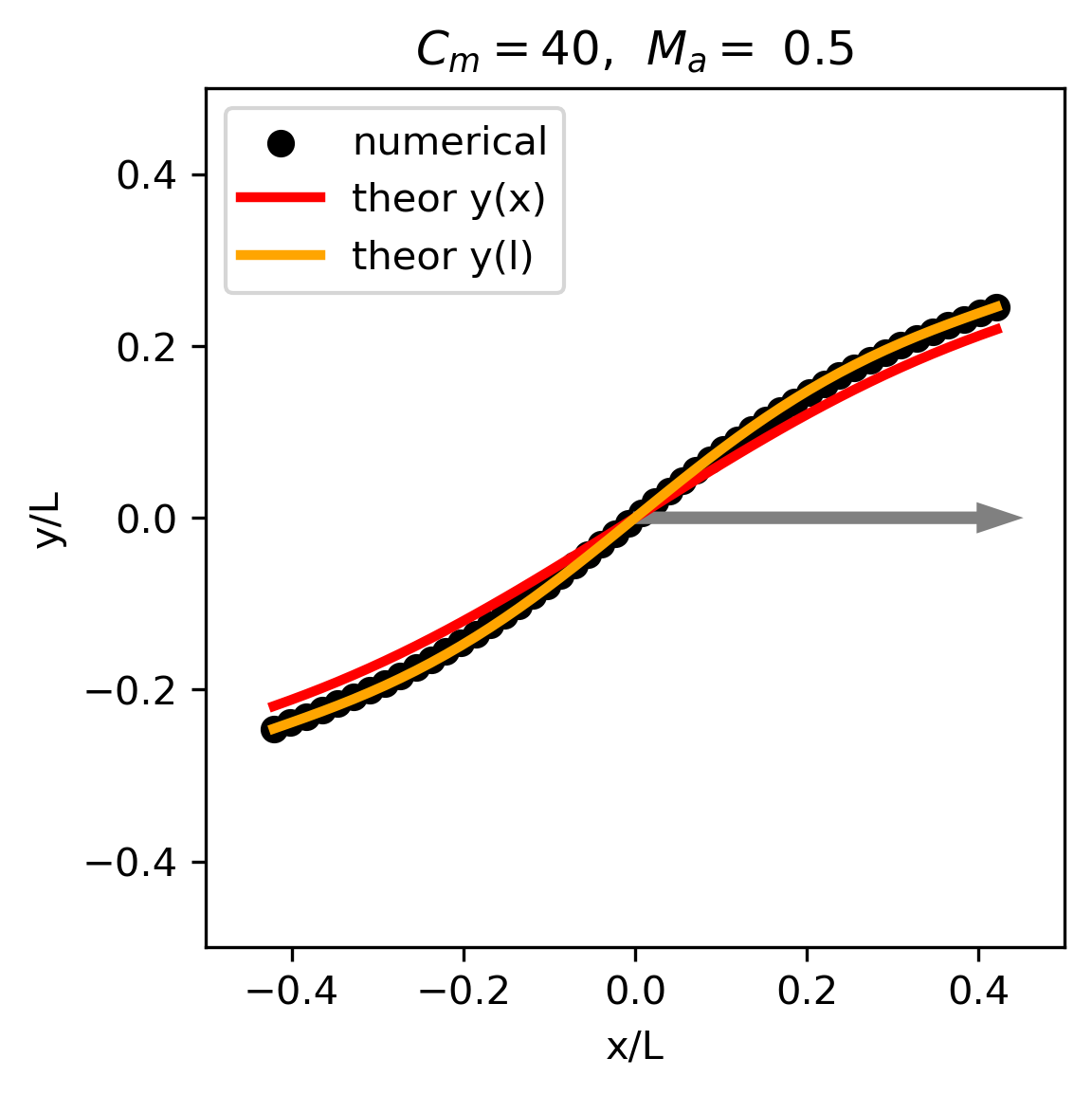}
    \includegraphics[width=0.32\textwidth]{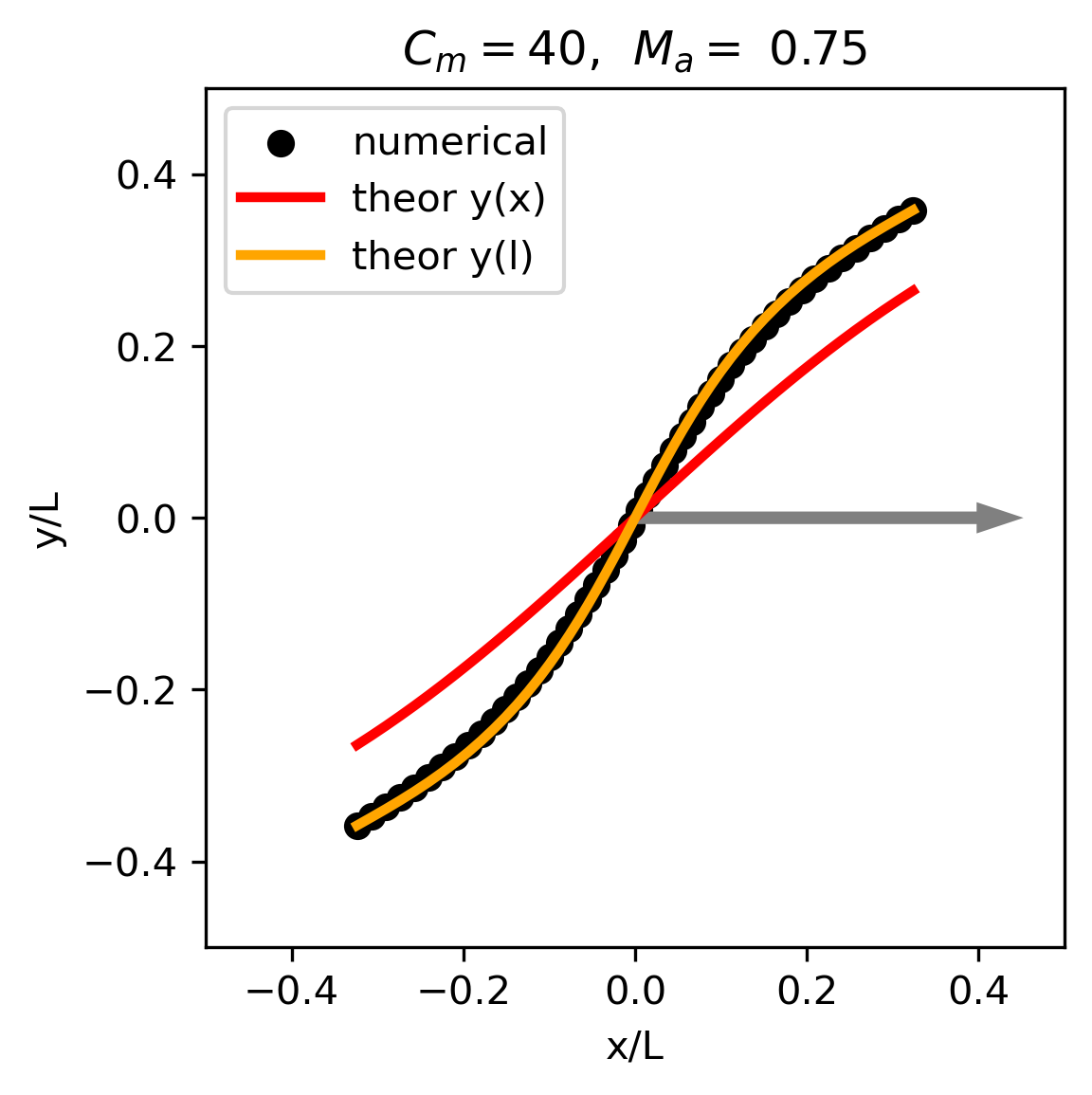}
    \includegraphics[width=0.32\textwidth]{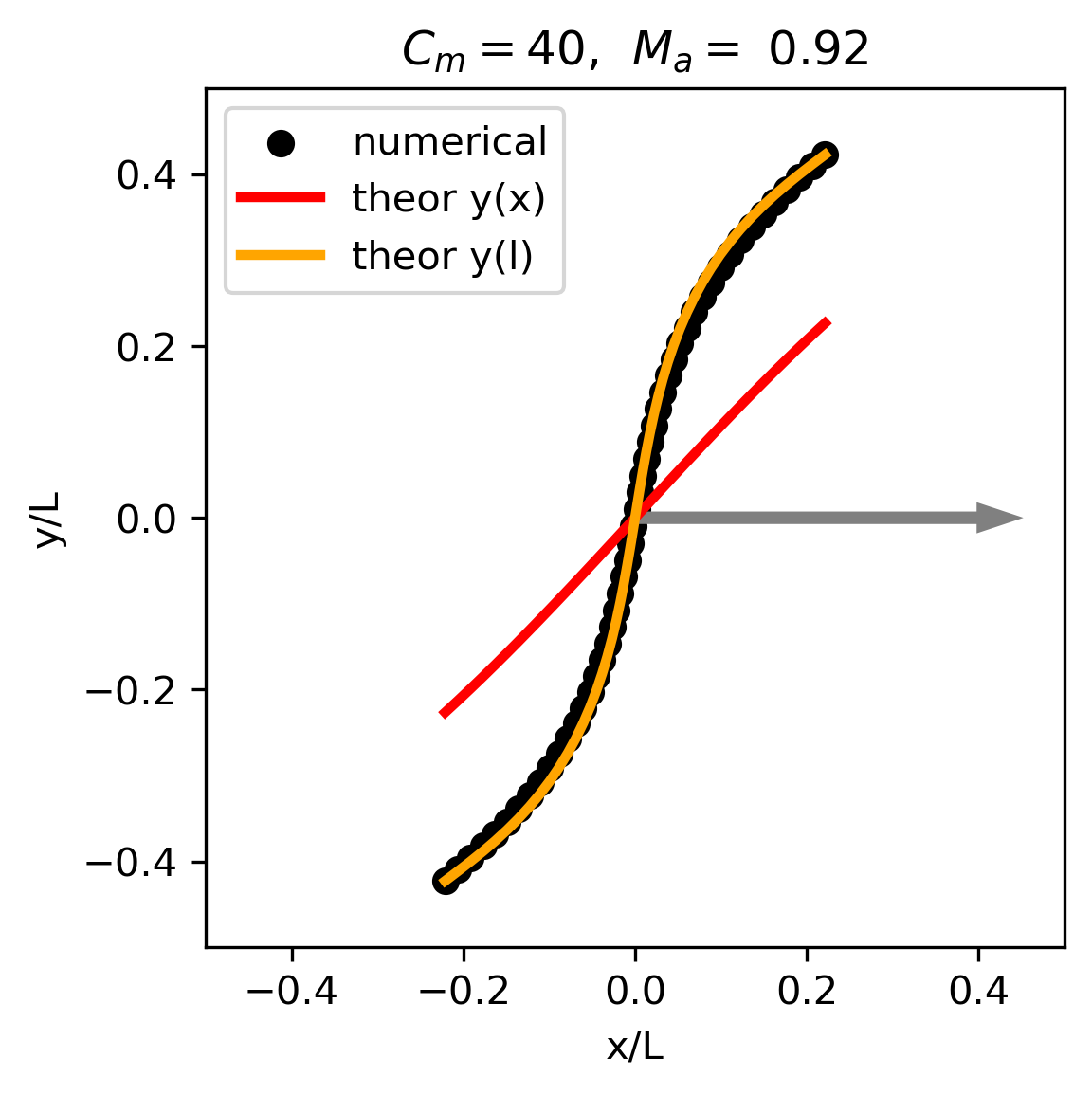}
    \includegraphics[width=0.32\textwidth]{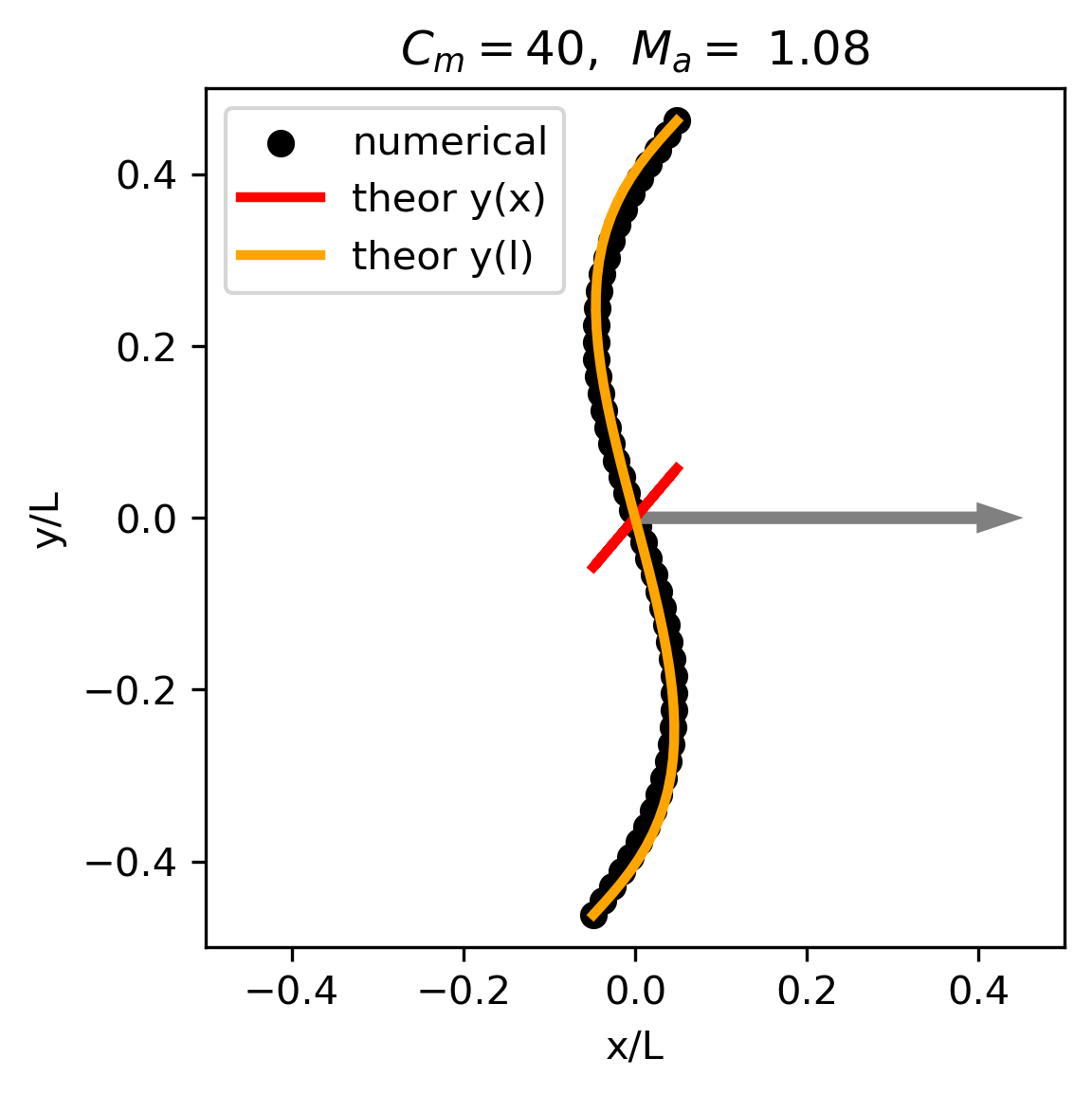}
    \caption{Equilibrium shapes of a rotating ferromagnetic filament for $Cm=40$ and for different $M_a$. The drag coefficient ratio is $\zeta_\perp/\zeta_\parallel=2$ in the numerical solution. The black dots show the numerical solution of the full problem, the red line shows the asymptotic solution (eq. \eqref{eq:dim_shape}) valid for small $M_a$, the yellow line shows the corrected asymptotic solution (eq. \eqref{eq:dim_shape_l}). The gray arrow shows the magnetic field direction.}
    \label{fig:simulation_shapes}
\end{figure}

We hope that this correction will improve the applicability range of the solution. 
Indeed looking at figure \ref{fig:simulation_shapes} we see that for small $M_a$ and deviations from the magnetic field direction, the both eqs. \eqref{eq:dim_shape} and \eqref{eq:dim_shape_l} well coincide with the full numerical solution.
As $M_a$ increases the corrected eq. \eqref{eq:dim_shape_l} follows the numerical shape much more closely, however, even it starts to noticeably deviate for $M_a>0.5$.
This is of course expected since in the derivation only the perpendicular drag coefficient $\zeta_\perp$ is used, but for larger deformations of the shape, the parallel drag coefficient $\zeta_\parallel$ starts to play a role.

\section{Procedure to determine the filament parameters}

\begin{figure}[h]
    \centering
    \includegraphics[width=0.49\textwidth]{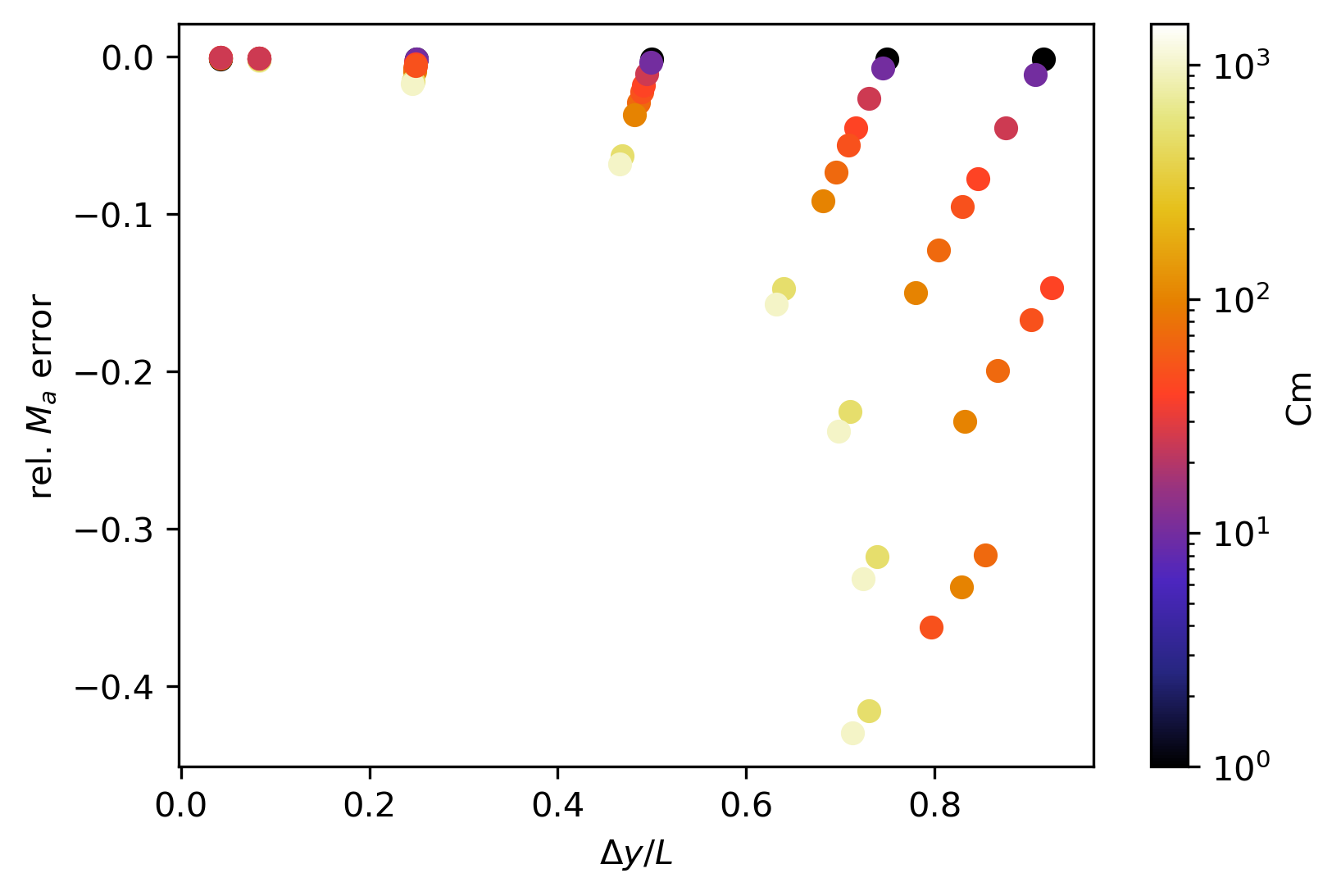}
    \includegraphics[width=0.49\textwidth]{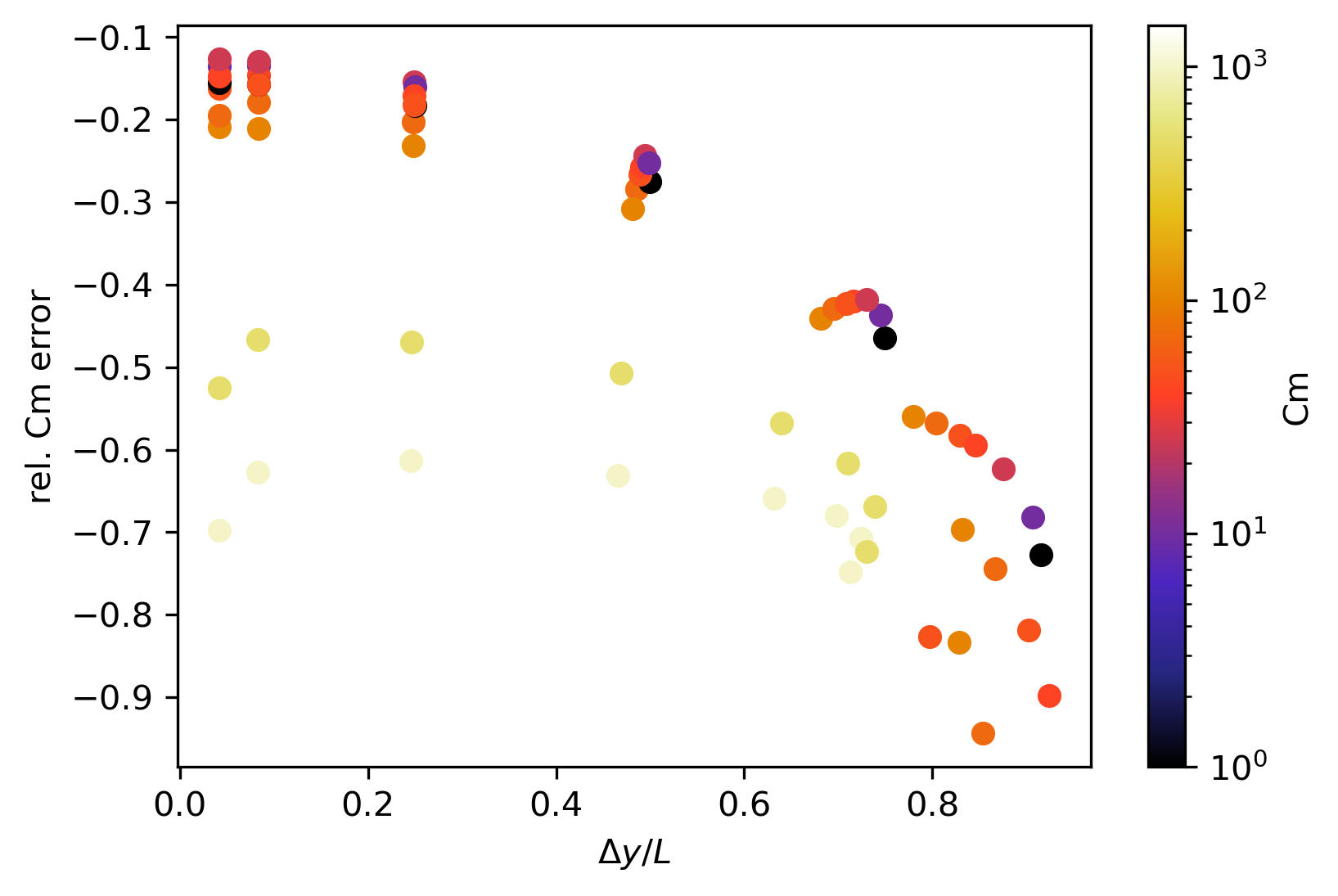}
    \caption{The relative error of the fitted $M_a$ (left) and $Cm$ (right) as a function of the tip deviation from the magnetic field direction $\Delta y/L$. The value of the magnetoelastic number $Cm$ used in the numerical equilibrium shapes is given by the color. }
    \label{fig:fit_error}
\end{figure}

The decoupling of the $M_a$ and $Cm$ effect on the shape (one determines the deviation from the field, while the other determines the shape) inspires us to propose the following procedure to determine them for an experimental filament. 
\begin{enumerate}
    \item Find the tips of the filament, and using eq. \eqref{eq:Ma_eq_l}, determine the Mason number $M_a =\Delta y / L$.
    \item Plug the found $M_a$ in eq. \eqref{eq:dim_shape_l} and determine $Cm$ by varying it until it best describes the filament's shape.
    \item Once the magnetic moment per unit length $M$ is known, calculate the bending modulus $A_B=\mu_0 H M L^2 / Cm$.
\end{enumerate}
Using this procedure, we determined the magnetoelastic $Cm$ and Mason $M_a$ numbers from the numerical equilibrium shapes.
The relative error $(fitted - true)/true$ of the parameters is shown in figure \ref{fig:fit_error}.

The determined $M_a$ is accurate for relatively large deviations from the magnetic field direction. 
The error in $M_a$ is less than $10\%$ for deviations of up to $\Delta y/L\approx 0.5$ from the magnetic field direction (which corresponds to roughly $30^\circ$ between the filament and the field). 
The deviation from the magnetic field direction can be experimentally controlled by the rotation frequency. 
One should choose a low enough frequency such that the deviations are small, but the shape is still visually discernible.
As expected, the equation \eqref{eq:Ma_eq_l} gives very accurate $M_a$ values for small $Cm$, which corresponds to nearly rigid rod. 
However, increasing $Cm$ values leads to an underestimate in $M_a$.

The error in the estimated $Cm$ is noticeably larger, and is dependent on $Cm$ itself.
The method is most accurate for $Cm\approx38$. 
As can be seen in figure \ref{fig:shape}, this is because for this $Cm$, the shape lies in between the two extreme configurations.
To minimize the error one should find the magnetic field value such that the experimentally observed shape is between the extremes of a straight rod and an S shape, whose tips align with the magnetic field.
Additionally, the deviation from the magnetic field direction should be less than $\approx20^\circ$ and $Cm\approx 10...70$ to have the relative error of $15...20 \%$.
Interestingly, the error seems to be systematic - the method underestimates the true $Cm$ value. 
We can therefore increase the fitted $Cm$ by $\approx 15\%$ to get a more accurate result.

To conclude the section let us examine a particular experimental observation (figure 1 (a) in ref. \cite{zaben_deformation_2020-1}, whose data are archived in ref. \cite{zaben_deformation_2020}).
We used the procedure outlined in the beginning of this section to estimate the $M_a$ and $Cm$ numbers.
Experimental filament's shape is taken from the center coordinates of the beads that make up the filament.
From the centers of the first and last bead of the filament we get 
$M_a=0.31$, and the shape fit then gives us $Cm=40\pm15$, which we can increase by $\approx 15\%$ (to offset the systematic underestimate as seen in figure \ref{fig:fit_error}) to obtain $Cm=46$.
From magnetization measurements it was found that these beads possess a magnetic moment of $m=1.4\cdot 10^{-13} A\cdot m^2$ \cite{erglis_three_2010}.
Dividing by their typical diameter $d=4.26 \mu m$ gives us the linear magnetization of the filament $M=3.3\cdot 10^{-8}A\cdot m$.
The length of the filament in the experiment is $L=67.4 \mu m$, and the magnetic field is $\mu_0 H = 0.86 mT$.
This allows us to determine the bending modulus $A_B=\mu_0 H M L^2 / Cm=2.8\cdot 10^{-21} J\cdot m$.
This result falls in the middle of the values shown in table \ref{tab:measured_AB}.
Finally, for comparison the effective bending elasticity that arises just from the interaction between magnetic dipoles in the chain is two orders of magnitude smaller $A_B^{mag}=\mu_0 M^2 (\zeta(3)+1/6)/(18\pi)=3.3\cdot10^{-23} J\cdot m$ \cite{vella_magneto-elastica_2014}, where $\zeta(n)$ is the Riemann zeta function.
This confirms that the bending stiffness mostly comes from the DNA linkers between the beads.

\section{Conclusions}

The equilibrium shape of a ferromagnetic filament in a rotating field contains the information about the filament's properties.
In particular, the tip positions relative to the magnetic field direction encode the value of the Mason number $M_a$ - the ratio of magnetic to viscous forces.
The shape that connects the tips is only dependent on the magnetoelastic number $Cm$ - the ratio of magnetic to elastic forces. 
For small values of $Cm$ the filament takes up a straight shape like a rigid magnetic rod, while for a large $Cm$ the filament's tips bend in the direction of the magnetic field, resulting in an S-like shape.
This allows us to determine $Cm$ just by visually observing the shape of the filament.
Once we know the magnetization of the filament, we can also determine the bending modulus $A_B$.

\section*{Declaration of Competing Interest}
The authors declare that they have no known competing financial interests or personal relationships that could have appeared to influence the work reported in this paper.

\section*{Acknowledgements}
Authors acknowledge the funding by the Latvian Council of Science, project A4Mswim, project No. lzp-2021/1-0470.

\printbibliography

\end{document}